\documentclass[twocolumn]{aastex62}
\usepackage{amsmath}
\usepackage{booktabs}
\usepackage{fontawesome}
\usepackage{soul, CJK}
\usepackage[utf8]{inputenc}
\usepackage{xspace}
\DeclareFixedFont{\ttb}{T1}{txtt}{bx}{n}{9} % for bold
\DeclareFixedFont{\ttm}{T1}{txtt}{m}{n}{9}  % for normal
\usepackage{color}
\definecolor{deepblue}{rgb}{0,0,0.5}
\definecolor{deepred}{rgb}{0.6,0,0}
\definecolor{deepgreen}{rgb}{0,0.5,0}
\usepackage{listings}
\newcommand\pythonstyle{\lstset{
language=Python,
basicstyle=\ttm,
otherkeywords={self},             % Add keywords here
keywordstyle=\ttb\color{deepblue},
emph={MyClass,__init__},          % Custom highlighting
emphstyle=\ttb\color{deepred},    % Custom highlighting style
stringstyle=\color{deepgreen},
frame=tb,                         % Any extra options here
showstringspaces=false            % 
}}
\lstnewenvironment{pyt}[1][]
{
\pythonstyle
\lstset{#1}
}
{}
\newcommand{\code}[1]{{\texttt{#1}}}

\newcommand{\Python}{\code{Python}\xspace}
\newcommand{\deepcr}{\code{deepCR}\xspace}
\newcommand{\deepCR}{\code{deepCR}\xspace}
\newcommand{\LACosmic}{\code{LACosmic}\xspace}

\graphicspath{{./}{figures/}}

\shorttitle{\deepCR: Cosmic Ray Rejection with Deep Learning}
\shortauthors{Zhang \& Bloom}
\begin{document}

\title{\deepCR: Cosmic Ray Rejection with Deep Learning}

\correspondingauthor{Keming Zhang}
\email{kemingz@berkeley.edu}

\author[0000-0002-9870-5695]{Keming Zhang \begin{CJK*}{UTF8}{gkai}(张可名)\end{CJK*}}
\affil{Department of Astronomy, University of California, Berkeley, CA 94720-3411, USA}

\author[0000-0002-7777-216X]{Joshua S. Bloom}
\affil{Department of Astronomy, University of California, Berkeley, CA 94720-3411, USA}
\affil{Lawrence Berkeley National Laboratory, 1 Cyclotron Road, MS 50B-4206, Berkeley, CA 94720, USA}



\begin{abstract}
Cosmic ray (CR) identification and replacement are critical components of imaging and spectroscopic reduction pipelines involving solid-state detectors. We present \deepCR, a deep learning based framework for CR identification and subsequent image inpainting based on the predicted CR mask. To demonstrate the effectiveness of this framework, we train and evaluate models on Hubble Space Telescope ACS/WFC images of sparse extragalactic fields, globular clusters, and resolved galaxies. We demonstrate that at a false positive rate of 0.5\%, \deepCR achieves close to 100\% detection rates in both extragalactic and globular cluster fields, and 91\% in resolved galaxy fields, which is a significant improvement over the current state-of-the-art method \LACosmic. Compared to a multicore CPU implementation of \LACosmic, \deepCR CR mask predictions run up to 6.5 times faster on CPU and 90 times faster on a single GPU. For image inpainting, the mean squared errors of \deepCR predictions are 20 times lower in globular cluster fields, 5 times lower in resolved galaxy fields, and 2.5 times lower in extragalactic fields, compared to the best performing non-neural technique tested. We present our framework and the trained models as an open-source \Python project \href{https://github.com/profjsb/deepCR}{\faGithub}, with a simple-to-use API.
To facilitate reproducibility of the results we also provide a benchmarking codebase \href{https://github.com/kmzzhang/deepCR-paper}{\faGithub}. 
\end{abstract}



\section{Introduction}
Charged particles passing through solid state detectors, such as charged coupled devices (CCDs), can corrupt astronomical imaging and spectroscopy by creating electron-hole pairs which in turn lead to excess charge in the affected pixels. Colloquially called ``cosmic rays'' (CRs), such charged particles can be terrestrial, instrumental, and cosmic in origin. Images taken with space-based facilities, such as the Hubble Space telescope (HST), are strongly prone to CR contamination, with the main culprit being electrons and protons trapped in Earth's radiation belt that originated in the Solar wind. The majority of CRs which affect CCDs in ground-based observations are secondary muons from high-energy showers, though radioactivity from instrument optics and dewar materials also contribute \citep{Groom2004-vq,fn95}.

Cosmic ray artifacts must be identified and either masked or replaced before further analysis can be done on such images. It is straightforward to identify these artifacts when multiple exposures of the same field are taken. In such cases, a median image could be calculated from aligned single exposures, effectively creating a CR-free image. Each one of the exposures is then compared with the median image to identify the CRs (cf.\ \citealt{Windhorst1994-ps}; see also \citealt{desai_detection_2016}).

However, when CCD read-out times are non-negligible, or when sources of interest are transient or variable, CR rejection with multiple exposures can be sub-optimal or infeasible. These cases require methods that find CRs in single exposures, such as linear filtering \citep{rhoads_cosmic-ray_2000}, median filtering (IRAF \code{xzap}), Laplacian edge detection (\citealt{van_dokkum_cosmic-ray_2001}; \LACosmic), and histogram analysis \citep{pych_fast_2004}, among others.  \cite{farage_evaluation_2005} evaluated the aforementioned methods and found \LACosmic to have the highest overall performance. \LACosmic identifies CRs by performing Laplacian transformations on upsampled input images to find CR edges, which tend to be much sharper than those of astronomical objects because cosmic ray artifacts are not smeared out by the atmosphere or telescope optics. Since edges only lie on the CR boundary, \LACosmic must be run with multiple iterations to mask larger CRs, with each successive step progressively identifying and replacing CR artifacts from the outermost pixels. Additionally, while Laplacian edge detection can be effective on well-sampled images, it can include many false detections when the point spread function (PSF) is critically sampled and becomes as sharp as CR artifacts, as in the case of HST images. \LACosmic tackles this by using a symmetry criteria (``fine structure image'') to distinguish arbitrarily shaped CRs from symmetrical, PSF spread-out point sources. The ``fine structure image'' is created by first convolving the image with a $3\times3$ median filter; for variations on scales of 2--3 pixels, this filter extracts a higher response for symmetric structure compared to non-symmetric structure. Larger scale variations are removed from the ``fine structure image'' by subtracting a convolution of the previous convolution with a $7\times7$ median filter. The Laplacian image is then divided by this ``fine structure image'' and compared against a threshold parameter, \texttt{objlim}, to exclude marginally sampled, symmetrical astrophysical sources from the CR mask.

In addition to the methods mentioned above, machine learning (ML) techniques for which classification rules are learned from labeled training data, have also been previously applied to CR detection; this includes neural networks \citep{1991ESOC...38...51M}, k-nearest neighbors, and decision trees \citep{Salzberg1995-cb}. However, these techniques have generally under-performed \LACosmic. In this paper, we present a deep learning based method for cosmic ray identification and replacement, which leverages recent progress in deep learning and computer vision. Central to these two fields are convolutional neural networks (CNNs; cf.\ \citealt{lecun_deep_2015}) which, in contrast to kernel convolution in \LACosmic, allow for kernels to be learned through back-propagation instead of being user-specified. In addition, compared with single-layer convolution, CNNs convolve feature maps recursively to potentially large depths, which allow semantic features to be extracted, in addition to low-level pixel features such as the location of edges. Recent developments in CNN architecture have advanced not only image classification \citep{deng_imagenet:_2009}, but also image segmentation (e.g., \citealt{shelhamer_fully_2017}), which refers to the process of labeling each pixel as belonging to a set of categories, and image inpainting, which refers to the process of predicting missing or corrupted parts of an image (e.g., \citealt{lehtinen_noise2noise:_2018}). In the context of CR rejection, image segmentation is a binary classification between cosmic ray artifact and non-artifact, while replacing pixel values for CR artifacts is essentially image inpainting.

Our primary interest in this paper is to develop a deep learning based CR detection and replacement method that is both more robust than existing techniques, and at least as fast to run. We motivate and define the architecture of our model in Section \ref{sec:model}, after which we discuss dataset construction in Section \ref{sec:data}. In Section \ref{sec:results}, we benchmark the performance of our model against baseline methods. A discussion of the results can be found in Section \ref{sec:summary}. Our framework and trained models are offered as an open-source project in \Python with a simple-to-use API\footnote{\url{https://github.com/profjsb/deepCR}}.

\begin{figure*}
 \centering
 \includegraphics[width=\textwidth]{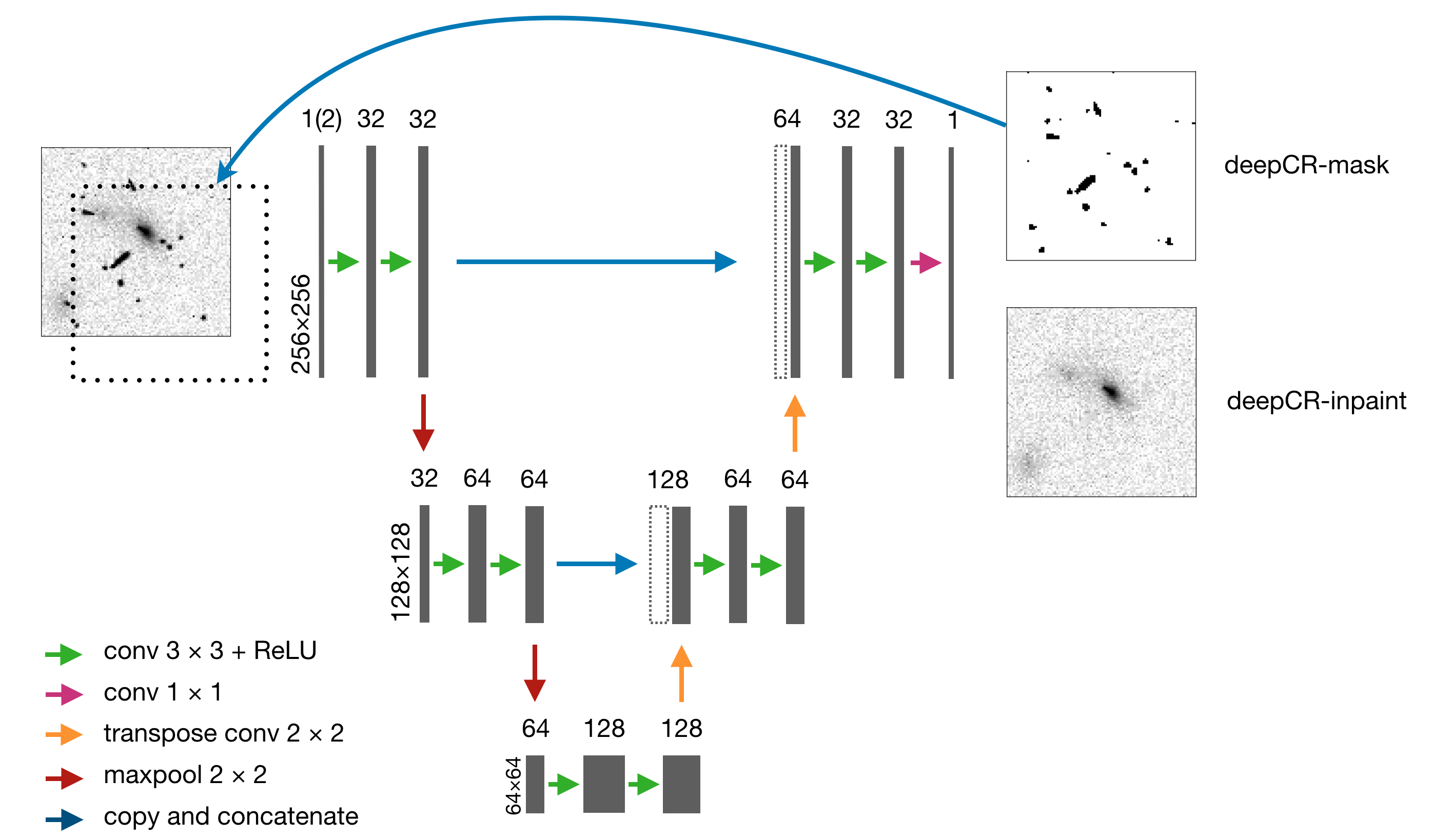}
 \caption{Neural network architecture of \deepCR. Feature maps are represented by gray boxes while the number of channels and feature map dimensions are indicated on the top of and to the left of each feature map, respectively. Different computational operations are marked in the legend to the lower left. ReLU stands for the rectifier activation function: $f(x)=\max(0,x)$. Unfilled boxes to the right of blue arrows represent feature maps directly copied from the left, which are to be concatenated with the adjacent feature map. To apply the inpainting model, the predicted mask (dotted box at left) is concatenated with the original image as input.}
 \label{fig:network}
\end{figure*}

\section{Model Architecture}
\label{sec:model}

We formulate our framework as a sequence of two independent deep neural networks, \deepCR-mask and \deepCR-inpaint:  \deepCR-mask, given an input image, predicts a probabilistic map of each pixel being affected by CRs, and \deepCR-inpaint predicts what the pixel values would have been had they not been affected by CRs. The probability map predicted by \deepCR-mask is then turned into a binary CR mask, i.e., 1 denoting cosmic ray, and 0 denoting non-artifact, by setting a threshold. The input image is then concatenated with the binary mask to be fed into \deepCR-inpaint, which outputs an image with the pixel values under the inpainting mask as the predicted values. The masked pixel values in the input image are set to 0 before feeding into concatenation. \deepCR-mask and \deepCR-inpaint are independently trained and can be used separately at application time. This means that \deepCR-inpaint, in principle, could take as input any inpainting mask so long as the unmasked pixels are free from CR artifacts.

The basic architecture of both \deepCR-mask and \deepCR-inpaint is a modification of the UNet \citep{ronneberger_u-net:_2015}, which is an encoder-decoder CNN with skip connections between each depth of the encoder and decoder (Figure \ref{fig:network}). The encoder extracts lower level pixel features such as edges at the first convolution layers, while deeper layers in the encoder closer to the network bottleneck generate higher level semantic information, e.g., locations and properties of CRs and stars. Information passes from the encoder to the decoder via the bottleneck and skip connections. The decoder up-samples the encoded information back to an output image in the original resolution. The skip connections allow the decoder to have direct access to low-level image feature, and thus are important components of the network because they allow the decoder to know not only the approximate location of CR artifacts (high-level features), but also their exact boundaries (low-level features).

UNet was first formulated for biomedical image segmentation and further adapted to a wide range of segmentation tasks, including transient detection in astronomy \citep{sedaghat_effective_2018}. Since UNet is fully convolutional and convolutional layers are translation invariant, it is not restricted to input images of fixed dimension. However, pixels near the image boundary have less contextual information compared to the rest of the image; for some applications this may limit the predictive power of the UNet for the outermost pixels. For this reason, the original UNet discards the boundary pixel predictions which do not have full contextual information in their receptive field; this results in segmentation maps with reduced spatial dimension. The number of pixels to be removed is network-depth dependent, and for their specific architecture, 92 pixels are to be removed on each image edge. However, unlike biomedical image segmentation, CR identification and replacement do not require long-range context; information from the few neighboring pixels are usually sufficient. We therefore require \deepCR predictions (segmentation and inpainting) to have the same dimension as the input image, thereby explicitly training \deepCR to do inference with limited contextual information at the image boundary. We find model performance near the boundary to have negligible difference compared with the interior of the image, although loss is marginally higher as expected. Our modification to UNet thus allows astronomers to retain near-boundary data in CR-cleaned images which otherwise would have been discarded.

Training \deepCR-mask and \deepCR-inpaint requires a set of training data which consists of CR-affected images (\textit{X}) and ground truth CR masks (\textit{M}). Our training data is constructed from HST images and includes multiple exposures of the same fields that provide the basis for us to derive accurate ground truth CR masks by comparing each exposure with a median image. Construction of the dataset is discussed in Section \ref{sec:data}.

We use the binary cross-entropy loss in training \deepCR-mask,
\begin{equation}
	\mathcal{L}_{\rm F} = \mathbb{E}[{\rm M}\times \log(1-\rm F(X))+(1-{M})\times \log(F(X))],
    \label{eq1}
\end{equation}
where F refers to the \deepCR-mask and $\mathbb{E}$ is the expectation with respect to training data. This loss essentially penalizes the log probability of each pixel prediction being wrong. For \deepCR-inpaint, the mean squared error (MSE) loss is taken between the predicted pixels under the inpainting mask (\textit{$M_{I}$}) and the ground truth values. While at first glance this would require having ground truth pixel values of CR artifact pixels, which are available to use in the median image, such necessity is easily circumvented by generating alternative inpainting masks in regions of the image that are not contaminated by cosmic rays. In practice, we select and add up one or more CR masks of other image stamps in the dataset as the inpainting mask; this naturally allows for data augmentation (see Section \ref{sec:aug}). One difference between using the median image as ground truth and the input image itself is that the median image is less noisy. Although at first glace training on higher S/N targets could be advantageous, \cite{lehtinen_noise2noise:_2018} showed that training on noisy targets is comparable to training on clean targets both in terms of convergence speed and final performance. The inpainting loss is then formulated as,

\begin{equation}
    \mathcal{L}_{\rm G} = \mathbb{E}[(\rm G(\rm X, \rm M_{I}) \circ \rm M_{I} \circ (1-\rm M)- \rm X \circ \rm M_{I} \circ (1-\rm M))^2],
    \label{eq2}
\end{equation}
\noindent where $\circ$ denotes element-wise multiplication. Here, G refers to \deepCR-inpaint, $X$ the input image, $M$ the cosmic ray mask, and $M_{I}$ the inpainting mask. Element-wise multiplication by $(1-M)$ ensures that the loss is not computed for parts of the inpainting mask that overlap with CR artifacts. Loss is also not computed outside of the inpainting mask; otherwise \deepCR-inpaint is forced to also learn an identity function for those regions, which is both unnecessary and degrades model performance at a fixed model capacity.

In our experiments, we also considered using a mean absolute error loss (L1) weighted by the inverse of image noise (Poisson and read noise). Although this loss is statistically better grounded, it penalizes predictions of sky background pixels much more than pixels with actual astrophysical flux. Considering that background pixels are already heavily weighted by the sheer number, and that pixel values of stars and galaxies are much harder to accurately reconstruct, we opt for the well-behaving MSE error instead.

\begin{table}
  \caption{Hubble ACS/WFC F606W fields used in dataset construction. Test data is marked with an asterisk ($^*$).}

  \centering
  \begin{tabular}{llllll}
    \toprule
    Proposal  & Visit & Type  & $N_{\rm exp}$ & $t_{\rm ave}$ \\
    ID & Num. & & & [sec] \\
    \toprule    
    10420 &01&extragalactic field&3&660\\
    10420 &05&extragalactic field&3&660\\
    10420 &07&extragalactic field&3&660\\
    12213 &11&extragalactic field&6&418\\
    13671$^*$ &44&extragalactic field&3&400\\
    10775$^*$ &06&globular cluster&4&100\\
    10775 &07&globular cluster&4&140\\
    10775 &08&globular cluster&5&380\\
    10775$^*$ &09&globular cluster&5&350\\
    10260 &03&resolved galaxy&3&790\\
    10260 &04&resolved galaxy&3&790\\
    10260 &06&resolved galaxy&3&790\\
    10260 &08&resolved galaxy&3&790\\
    10584 &32&resolved galaxy&3&527\\
    10190$^*$ &28&resolved galaxy&4&540\\

    \toprule
  \end{tabular}
  \label{table:data}
\end{table}

\section{Data}
\label{sec:data}
We have constructed our training and testing datasets from HST ACS/WFC imaging data, using the {\sc astrodrizzle} \citep{astrodrizzle} data pipeline. The imaging data we used is listed in Table \ref{table:data}, and is divided into three categories: extragalactic field, globular cluster, and local group galaxies for which the stellar population is well resolved; the density of astronomical sources, which are also sources of confusion during CR identification, increases in that order. While we train simultaneously on the three categories of data, evaluation is done separately to examine the performance of each model on tasks of various difficulty. 

Each set of images consists of 3 to 6 single exposures. {\sc astrodrizzle} creates a CR-free image for each calibrated single-frame exposure (\texttt{*flc.fits}) by aligning every frame to sub-grid accuracy and calculating a median image, before ``blotting'' the median image back into the grid of each individual exposure. It then identifies CRs in each frame by comparing with both the median image and a derivative median image which accounts for discrepancies caused by alignment residuals. CRs are then identified from the difference image with two passes. A preliminary CR mask is first produced with a high signal-to-noise (S/N) threshold to minimize false detections. Pixels adjacent to the candidates found in the first pass, specified by a growing radius (set to 1 pixel), are examined with a lower S/N threshold to identify the dimmer peripherals of each CR. While the default S/N thresholds are 3.5 and 3 for the first and second pass, we opted for 5 and 1.5 because the default first threshold causes many false positives and the second threshold is not low enough for it to fully mask the dimmer peripherals of CRs.

To ensure good training behavior, we also created a bad pixel mask and a saturation mask for which we do not backpropagate through nor evaluate the models on. The bad pixel masks are derived from the data quality array in the \texttt{*flc.fits} files where we include all flags except the saturation flag. We create a conservative saturation mask by masking pixels brighter than \hbox{70000 e$^-$}, which we further expand with a $7\times7$ dilation kernel to make sure that the peripherals of blooming artifacts are covered.

Finally, we divide the images and masks into image stamps of $256\times256$ to facilitate batch training. Because each single frame image may be slightly offset from each other due to dithering, we discard the first 128 pixels at the image boundary to ensure that each image used has complete overlap with at least one other image. Each frame of exposure then yields 210 image stamps. This results in a training set of 8190 image stamps and a test set of 3360 image stamps. We further reserve 1638 image stamps from the training set as the validation set to monitor over-fitting during training. To expose any additional over-fitting, the test set images are chosen to have different target fields than the training images and a wide range of exposure times.

\begin{figure}
 \centering
 \includegraphics[width=\linewidth]{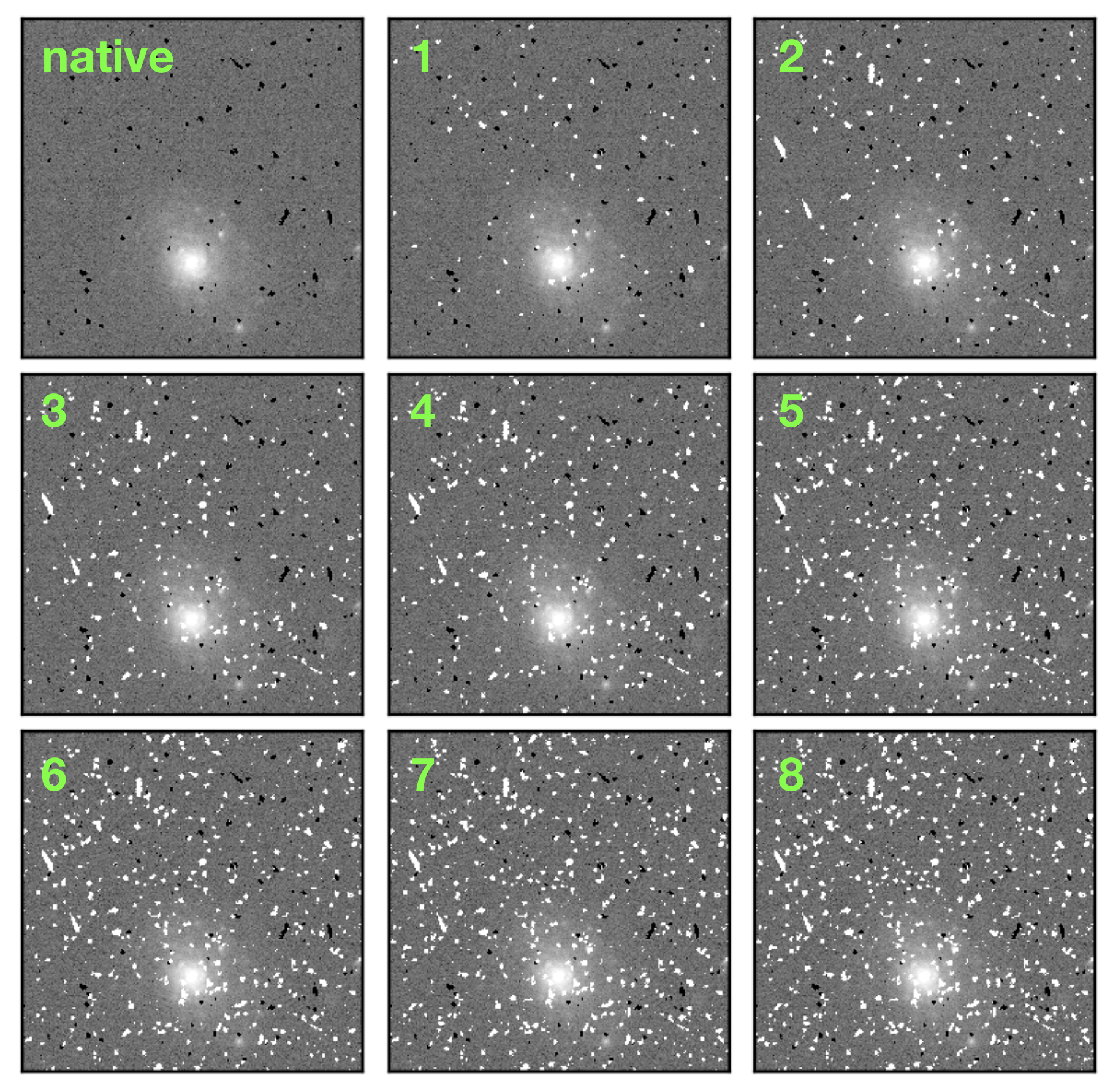}
 \caption{Data augmentation scheme for creating inpainting masks. Numbers indicate the number of CR masks sampled and added from the dataset to create the inpainting mask, as shown in white. The original CR artifacts are masked in black to distinguish them from the inpainting mask.}
 \label{fig:inpaint_aug}
\end{figure}

\begin{figure*}
 \centering
 \includegraphics[width=\textwidth]{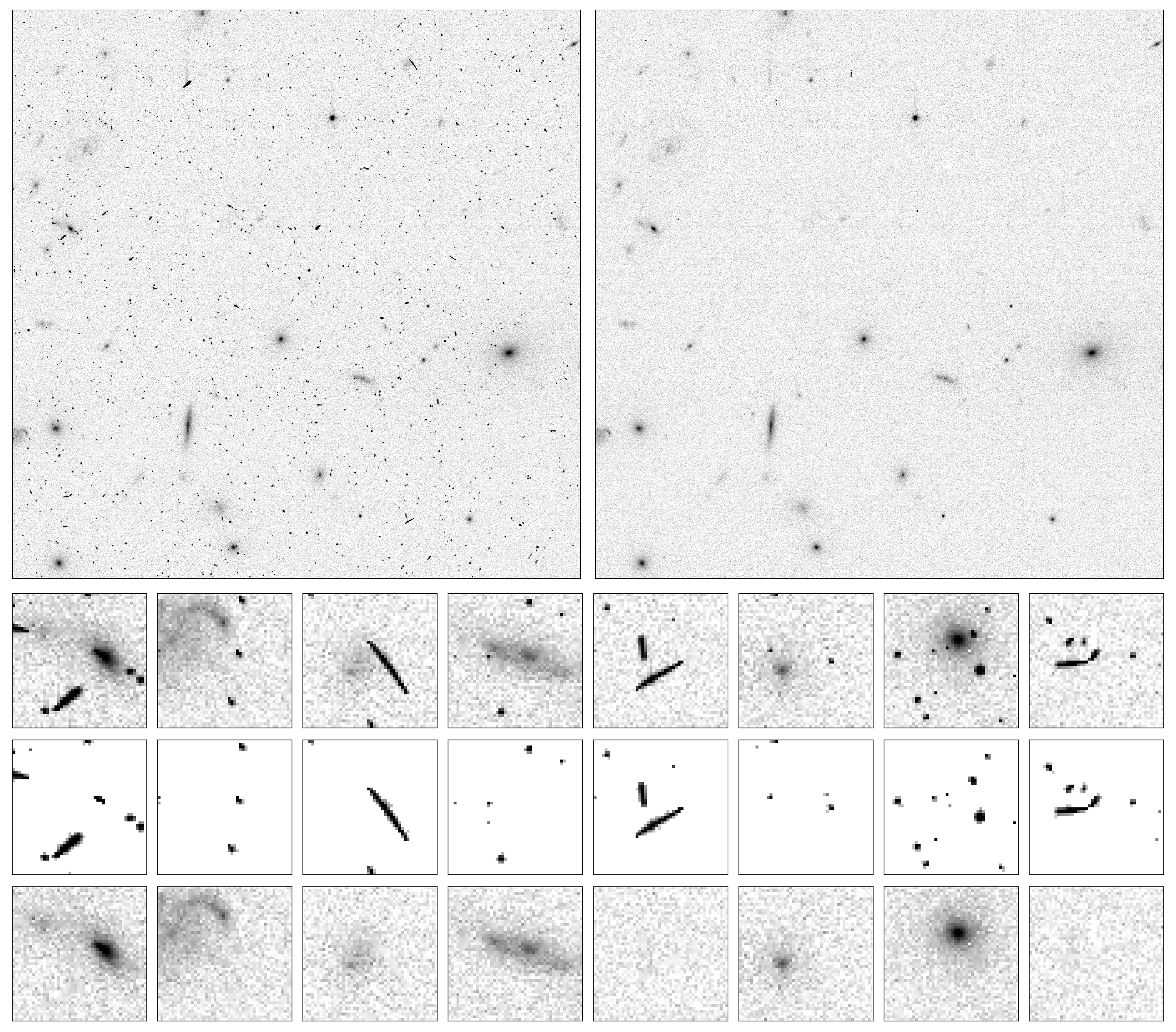}
 \caption{(top left) ACS/WFC image of galaxy cluster EMACSJ2316.6 (HST:13671-44). (top right) The same image with CRs removed and inpainted by \deepCR. (bottom) The top row shows details of some CR artifacts. The middle panel shows CRs as identified by \deepCR-mask-2-32. Bottom panel shows images with artifact pixels replaced by \deepCR-inpaint-3-32.}
 \label{fig:mask_example}
\end{figure*}

\subsection{Data Augmentation}
\label{sec:aug} 
During both training and evaluation, we create an inpainting mask for each image by sampling and adding 1--9 CR masks from the rest of the dataset. Our augmentation scheme allows \deepCR-inpaint to adapt to inpainting masks of varying density (Figure \ref{fig:inpaint_aug}); having denser inpainting masks also allows for more efficient training.
 
Additionally, we also augment the image sky background level in training both \deepCR-mask and \deepCR-inpaint to remedy for the limited and discrete exposure times in the training set. Different exposure times and sky background levels mainly change the contrast of CR artifacts (as well as astronomical objects) against the background, which can affect model prediction. We adjust the sky background level by adding up to 3 times and subtracting up to 0.9 times the original level. Since the original pixel value can be written as
\begin{equation}
	n = (f_{\rm star} + f_{\rm sky}) \cdot t_{\rm exp} + n_{\rm CR},
\end{equation}
where $n$ is in units of e$^-$ and flux ($f$) in units of e$^-$/s, the pixel value after augmentation is
\begin{equation}
	n^{\prime} = n + \alpha \cdot f_{\rm sky} \cdot t_{\rm exp} \\
    = \left(\frac{f_{\rm star}}{1+\alpha} + f_{\rm sky}\right) \cdot (1+\alpha)\cdot t_{\rm exp} + n_{\rm CR}.
\end{equation}

\begin{figure*}
 \centering
 \includegraphics[width=\textwidth]{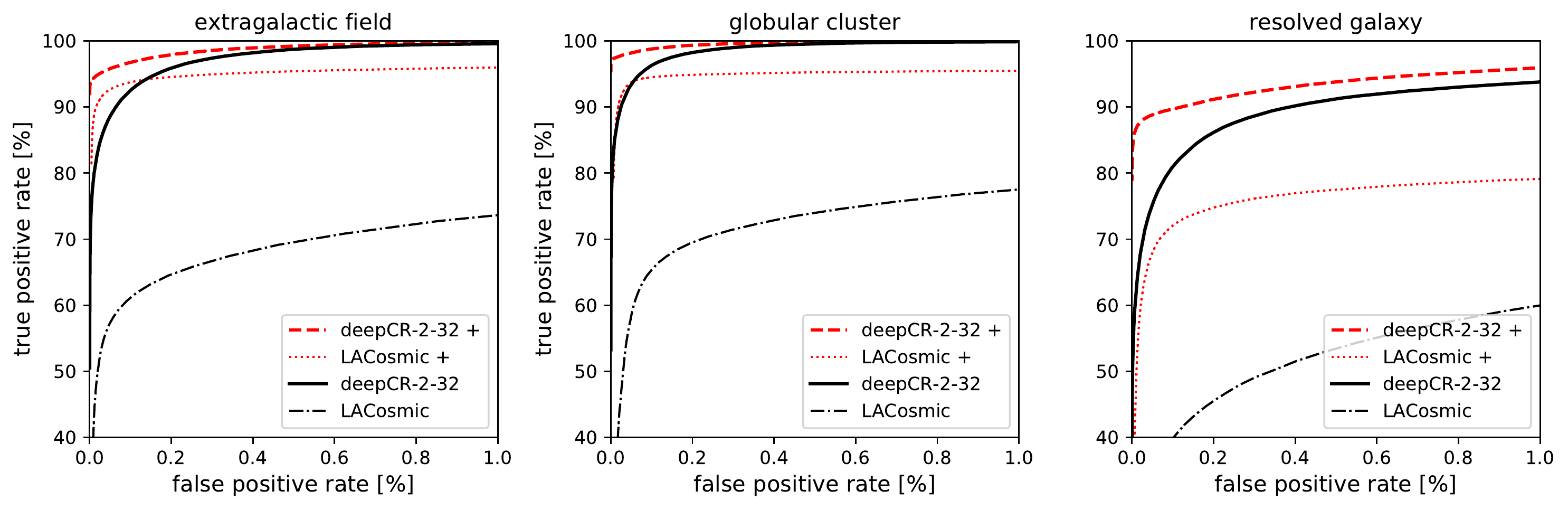}
 \caption{Receiver operating characteristic (ROC) curves of cosmic ray detection. Line shapes and the corresponding methods are indicated in the legends to the lower right. Different points on the ROC curves correspond to different thresholds adapted in \deepCR and \LACosmic. Red ROC curves (labeled with + in the legend) differ from black ones in that true positive rates are calculated from $3\times3$ kernel dilated CR mask predictions, with false positive rates kept the same at the non-dilated values. The red curves therefore do not penalize a model for not finding the exact cosmic ray shape matching the ground truth.}
 \label{fig:ROC}
\end{figure*}

\begin{table*}[t]
 \caption{Cosmic ray true detection rates (TPR) evaluated on test set images of extragalactic fields, globular clusters, and resolved galaxies, at fixed false positive rates (FPR) of 0.05\% and 0.5\%, shown in top and bottom tables respectively. The fixed FPRs are specified in the headers. TPRs in parentheses show the detection rates after mask dilation, as described in the main text. The last two columns show the runtime required to predict CR masks for one hundred $256\times256$ images, evaluated on either 4 cores of an Intel Xeon CPU, or 1 Nvidia Titan X GPU. The \deepCR runtimes are evaluated as the time to complete a forward pass through the trained model.}
 \centering
  \begin{tabular}{*{6}{c}}
 \toprule 
          & extragalactic field & globular cluster & resolved galaxy & \multicolumn{2}{c}{runtime [sec]}   \\
        Model &TPR (0.05\%)& TPR (0.05\%)& TPR (0.05\%)& CPU & GPU \\
 \midrule
\deepCR-2-4 &82.0\% (94.5\%) &83.9\% (97.3\%) &56.2\% (80.4\%) &\textbf{1.4} &\textbf{0.1}\\
\deepCR-2-32 &\textbf{88.5\%} (95.8\%) &\textbf{93.3\%} (98.1\%)&\textbf{75.2\%} (88.8\%)&7.9 &0.2\\
\LACosmic &57.3\% (92.6\%) &58.3\% (93.7\%) &33.8\% (67.9\%) &9.0 & n/a\\
 \midrule 
 \midrule 
 &TPR (0.5\%)& TPR (0.5\%)& TPR (0.5\%) \\
 \midrule 
\deepCR-2-4 &94.0\% (96.9\%) &96.2\% (98.7\%) &80.6\% (89.7\%) &\\
\deepCR-2-32 &\textbf{98.7\%} (99.2\%) &\textbf{99.5\%} (99.7\%)&\textbf{91.2\%} (93.7\%)&\\
\LACosmic &69.5\% (95.4\%) &73.9\% (95.2\%) &53.4\% (77.4\%) &\\
\bottomrule 
 \end{tabular}
 \label{table:mask}
\end{table*}

Thus, adding or subtracting a multiplicative of the sky level, i.e., $\alpha\cdot f_{\rm sky}$, is equivalent to simulating an exposure time of $(1+\alpha)\cdot t_{\rm exp}$, with flux from astronomical objects scaled down by $1+\alpha$, which is of minimal concern since astronomical fluxes already span orders of magnitude. Although alternatively one may simulate different exposure times by directly scaling the image with a multiplicative factor, doing so would lead to different CR statistics, whose contribution to the observed pixel value is independent of integration time. While our augmentation scheme inevitably changes the image noise properties, following similar arguments in \cite{lehtinen_noise2noise:_2018}, homogeneous noise in the training set should have little effect on model performance. In our experiments, we observed that a model trained with sky background augmentation performed significantly better on test data augmented in the same way, and equally well on native test data. On real data, \deepCR-mask trained with augmentation improves detection rates for the shortest exposure (100 seconds) test set images (HST \hbox{10775-06}) by $\sim2\%$ when images with exposure times less than 400 seconds are withheld from the training set. Including these short exposure data in our training set did not further improve our metrics. This shows that the sky background augmentation scheme is as helpful as having more data of different exposures. We therefore omitted short-exposure globular cluster data in training \deepCR-mask.

\section{Results}
\label{sec:results}
We label each variant of our network with two hyper-parameters, i.e., \deepCR-3-32 would be a depth-3 network with 32 channels at the convolution layer (same as Figure \ref{fig:network}). The labels of mask and inpaint are omitted whenever context is evident. For \deepCR-mask, we trained two different variants: \deepCR-2-4 and \deepCR-2-32. We also tried architectures as large as \deepCR-4-64, but did not see significant improvements from \deepCR-2-32. For \deepCR-inpaint, we trained and evaluated on \deepCR-2-32 and \deepCR-3-32. Details on training can be found in Appendix \ref{sec:training}. We benchmarked network performances against baseline models, and describe the results as follows.

\begin{table*}
 \caption{Image inpainting metrics of \deepCR-inpaint as compared to those of baseline models. MSE is the mean squared error of 9600 generated test image stamps. The last two columns show runtime to inpaint one hundred $256\times 256$ image stamps, evaluated either on 4 cores of an Intel Xeon CPU, or one Nvidia Titan X GPU. \deepCR runtimes are evaluated as the time to complete a forward pass through the trained model.}
 \centering
  \begin{tabular}{llllll} 
 \toprule 
         & extragalactic    & globular cluster & resolved galaxy & \multicolumn{2}{c}{runtime [sec]}\\
        Model & MSE & MSE& MSE & CPU & GPU  \\
 \midrule
  \deepCR-2-32&\textbf{0.012} & 0.034 & 0.503 & 7.5 & \textbf{0.2}  \\
  \deepCR-3-32&\textbf{0.012} &\textbf{0.033} &\textbf{0.479} & 12.7 & 0.3  \\
  medmask&0.105 & 1.511 & 5.946 & \textbf{1.0}\footnote{Estimation from a single threaded implementation, which took 8.0 seconds on average.} & n/a \\
  biharmonic&0.041 & 0.669 & 2.578 & 109.5 & n/a \\
\bottomrule 
 \end{tabular}
 \label{table:inpaint}
\end{table*}

\subsection{Cosmic Ray Identification: \deepCR-mask}
We evaluate \deepCR-mask with Receiver Operating Characteristic curves (ROC curves), against the baseline model of \LACosmic. ROC curves show true positive rates (fraction of CR artifact pixel identified) as a function of false positive rates (fraction of pixels mistaken as CRs) and are plotted by dialing the threshold parameter of a given model. For \LACosmic, an additional parameter \texttt{objlim} which controls the symmetry discriminant (the ``fine structure image'') needs to be tuned for its optimal performance. We experimented with a range of values, and found \texttt{objlim}=2, 3.5, and 5 to be the optimal values for images of extragalactic fields, globular cluster, and resolved galaxy, respectively. We note that \texttt{objlim}=2 and 3.5 are smaller than the recommended value of 4--5 for HST WFPC in the original \LACosmic documentation, but smaller \texttt{objlim} value allows fewer CRs to be discounted on the basis of symmetry, which is advantageous when few unresolved sources of confusion are present in the image.

Figure \ref{fig:ROC} shows ROC curves of \deepCR-2-32 (hereafter \deepCR) and \LACosmic evaluated on the three categories of data in the test set. As seen in the black curves in Figure \ref{fig:ROC}, at fixed false positive rates (FPR), \deepCR is able to achieve much higher true positive rates (TPR) compared with \LACosmic for all three fields. While \deepCR is able to achieve nearly 100\% TPRs for both extragalactic fields and globular cluster fields and $>90\%$ for resolved galaxy fields, the detection rates of \LACosmic is consistently below 80\% for the same FPR range.

\begin{figure}
 \centering
 \includegraphics[width=\linewidth]{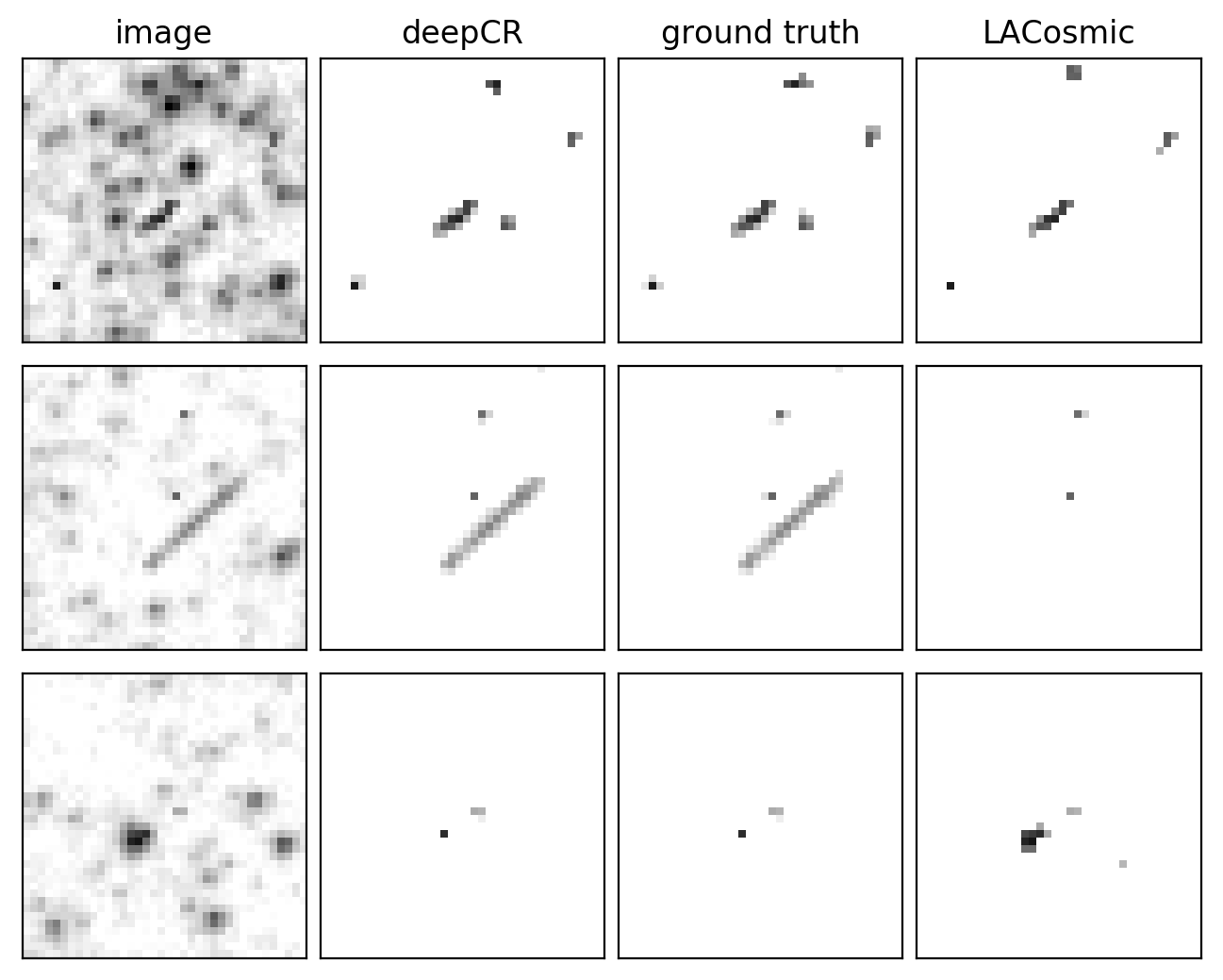}
 \caption{Comparison of \deepCR-mask-2-32 and \LACosmic in 3 resolved galaxy field examples. Columns from left to right are: the input image, \deepCR predicted CR, ground truth CR, and \LACosmic predicted CR. Threshold for \deepCR  is set to 0.5 while \LACosmic uses the following parameters: \texttt{objlim=5}, \texttt{sigclip=10}, where the threshold \texttt{sigclip}, is chosen to balance true detection and false detection. \deepCR prediction is about identical to ground truth in all cases. Rows 1 and 2 show the tendency of \LACosmic to miss either the peripheral or the entirety of larger CRs. Rows 1 and 3 show cases where \LACosmic misidentifies stars as cosmic rays.}
 \label{fig:mask_compare}
\end{figure}

However, to make a conclusion from here would potentially be unfair to \LACosmic, because imaging noise makes it impossible to create objective ground truth CR masks, and the characteristic way by which \texttt{drizzlepac} creates CR masks could be learned by \deepCR during training but not \LACosmic. In particular, we adapted relatively liberal thresholds in the two-step CR identification procedure in \texttt{drizzlepac} to minimize residual CR flux in the image, but \LACosmic tend to ignore the peripheral pixels of larger CRs. In practice, users may choose to expand the predicted mask for a more conservative treatment of CRs. Therefore, to remove any aforementioned advantage \deepCR may have gained, we re-calculated true positive rates for both models on $3\times3$ kernel dilated (expanded) mask predictions, while keeping FPRs fixed at values evaluated on non-dilated masks. $3\times3$ kernel dilation essentially masks all 9 pixels in any $3\times3$ region as long as the center pixel is identified as cosmic ray in the original mask. The re-calculated ROC curves, shown in red in Figure\ \ref{fig:ROC}, therefore do not penalize a model for not finding the exact CR shape matching the ground truth, thereby removing any advantage \deepCR may have gained.

As expected, mask dilation increases true detection rates drastically for \LACosmic, also slightly improving ours at the very low-end of FPRs. However, our model still retains a significant advantage in all three types of fields. \LACosmic detection rates in extragalactic fields and globular cluster fields are increased to up to 95\% where the ROC curve plateaus; 5\% of CRs are consistently missed regardless of detection threshold. In resolved galaxy fields, our model still outperforms \LACosmic by $>20\%$ in TPR. The advantages are more significant at very low false detection rates. Figure \ref{fig:mask_compare} shows examples of mask prediction by \deepCR and \LACosmic in resolved galaxy fields. As seen in the figure, the CR mask predicted by \deepCR is about identical to ground truth in all cases, while \LACosmic tends to miss larger CRs and is prone to false detections. Further quantitative comparison is presented in Table \ref{table:mask}, which also lists the runtime required to predict one hundred $256\times 256$ image stamps. We find that the smaller variant, \deepCR-2-4, runs around 90 times faster on one Nvidia Titan X GPU than a multicore CPU implementation of \LACosmic, and 6 times faster on CPU. The CPU implementation of each model is well parallelized and fully utilizes the CPU cores available (4 cores on an Intel Xeon CPU for our benchmarking).

\subsection{Replacing Masked Pixels: \deepCR-inpaint}
We evaluate \deepCR-inpaint against non-neural baseline models including biharmonic interpolation and masked median sampling (5$\times$5 filter; \texttt{medmask}) that is used by \LACosmic. Evaluation is done on 9600 image stamps randomly chosen from the test set, with different generated inpainting mask applied to each one of them. The masks are generated as described in Section \ref{sec:aug}, with the number of added masks for each image stamp fixed at 1 to preserve the native mask density in the test set. By fixing \texttt{numpy} random seeds, we make sure all models are evaluated on the same sequence of image stamps with the same inpainting masks applied.

Inpainting performances and runtimes are listed in Table \ref{table:inpaint}, which shows that \deepCR achieves MSEs 20 times lower in globular cluster fields, 5 times lower in resolved galaxy fields, and 2.5 times lower in extragalactic fields, compared to the best performing non-neural method we tested, biharmonic inpainting. Our model is also drastically faster than non-neural models on GPU. For the non-neural models, biharmonic is more accurate than masked median sampling, though at a cost of $100\times$ compute. Multicore implementation of median masked sampling remains the fastest method on CPU. Although \deepCR-inpaint is comparably accurate, we caution that further testing with astronomically meaningful metrics (such as photometric fidelity) is required, before the inpainted pixels can be used for directly for quantitative science tasks.

\subsection{Usage}
To accompany this paper we provide a \Python package called \deepCR \href{https://github.com/profjsb/deepCR}{\faGithub} which can be installed via {\sc pip}.  The models trained on HST ACS/WFC F606W data which are used in this paper are shipped along with the code. If \texttt{image} is a 2-dimensional \texttt{numpy} image array that contains CRs, the procedure to predict a CR mask (\texttt{mask}) and a clean image (\texttt{cleaned\_image}) is as follows:
\begin{pyt}
from deepCR import deepCR
mdl = deepCR(mask="ACS-WFC-F606W-2-32",
	     inpaint="ACS-WFC-F606W-3-32",
             device="GPU")
mask, cleaned_image = mdl.clean(image)
\end{pyt}

\deepCR is first instantiated with the specified model configuration, before applying to data. Non-neural inpainting could be used in place of \deepCR-inpaint by setting \texttt{inpaint="medmask"}.
We intend to host a ``model zoo'' for models learned on other instrumental configurations and welcome community engagement to generate such models and to improve the code.

\section{Summary}
\label{sec:summary}
We have presented a novel deep learning based approach for cosmic ray rejection, and showed that after proper training, it has the potential to outperform current methods in terms of both speed and accuracy in mask prediction and image inpainting. At a false positive rate of 0.5\%, \deepCR achieves close to 100\% detection rates in extragalactic and globular cluster fields, and 91\% in resolved galaxy fields, which is a significant improvement over current state-of-the-art, \LACosmic. Compared to a multicore CPU implementation of \LACosmic, \deepCR-mask runs up to $6.5\times$ faster on CPU and 90$\times$ faster on GPU. As for inpainting, mean squared errors of \deepCR-inpaint predictions are 20 times lower in globular cluster fields, 5 times lower in resolved galaxy fields, and 2.5 times lower in extragalactic fields, compared to biharmonic inpainting. The superior MSE performance of the \deepCR-inpaint over generic interpolation schemes is not surprising, given that our model is trained on the semantically constrained domain of astronomical images. To facilitate reproducibility of these results, we have released the benchmarking codebase\footnote{\url{https://github.com/kmzzhang/deepCR-paper}}.

While \LACosmic requires fine tuning of \texttt{objlim} for different instrumental setups to maximize CR detection accuracy and minimize false detection, our approach requires training on a set of CR-labeled images that are representative of new data expected from the data reduction pipeline adapting \deepCR. Since CNN models are based on pattern recognition learned from training data, new training data may be required if predict-time data differs significantly from training data. Our work has focused on HST ACS/WFC imaging in a single filter for which there was readily available training data across a variety of field types. However, we suggest that models trained on one particular filter of a detector will likely perform well on other filters of the same instrument, and more likely on filters of longer wavelengths, because with wider PSFs sources of confusion would have less resemblance to cosmic rays. Indeed, in our preliminary experiments, we blindly applied \deepCR-mask trained on the ACS/WFC F606W filter to ACS/WFC F814W data, and found comparable performance. As we have trained a single model to work across three different types of fields, it should also be possible to train a well-performing model on several different filters of the same instrument, with a similar sized training dataset.

Looking ahead, we speculate that it may be feasible to train a single model not only on different filters, but also on different detectors and telescopes, though this would certainly require larger capacity models and more sophisticated data pre-processing. We have not experimented with spectroscopic data nor ground-based data, but expect our approach to work as well. To facilitate the use of the \deepCR framework in real-world reduction pipelines, we have made our code with the ACS/WFC F606W trained models available as an open-source project, and we encourage the community to contribute by training additional models that allows \deepCR to be used in a wide range of detector configurations.

\software{%
    \texttt{astropy} \citep{2013A&A...558A..33A, Price-Whelan18},
    \texttt{astrodrizzle} \citep{astrodrizzle},
    \texttt{numpy} \citep{vdw11},
    \texttt{scipy} \citep{scipy},
    \texttt{matplotlib} \citep{2007CSE.....9...90H}, 
    \texttt{astroscrappy} \citep{curtis_mccully_2018_1482019},
    \texttt{pytorch} \citep{paszke2017automatic},
    \texttt{Jupyter} \citep{Kluyver:2016aa},
    \texttt{scikit-image} \citep{walt_scikit-image:_2014}.
}

\acknowledgements{It is a pleasure to thank Dan Weisz for helpful conversations and St\'efan van der Walt both for his insights on image inpainting and for comments on a draft of this manuscript. We thank the anonymous referee for their close reading of the original version and for the suggestions that led to improvements in the published paper. This work is supported by a Gordon and Betty Moore Foundation Data-Driven Discovery grant.}

\appendix

\section{Training procedures}
\label{sec:training}
We implemented \deepcr models in {\sc Pytorch 1.0} \citep{paszke_automatic_2017}. We follow the steps described below to train both \deepCR-mask and \deepCR-inpaint. The network is first trained for 20 epochs (40 for \deepCR-inpaint) in ``training mode'' for which batch normalization layers keep a running statistics of layer activations with a momentum of 0.005, and use training batch statistics for normalization. Following this initial training phase, the network is set to ``evaluation mode'' for which the running statistics are frozen and used in both forward and backward passes of batch normalization. This procedure ensures that batch normalization statistics used at training and test time are identical, which helps the network achieve lower loss. Given the large dynamic range of astronomical imaging data, we have found this technique to be essential.

\begin{figure*}
 \centering
 \includegraphics[width=\textwidth]{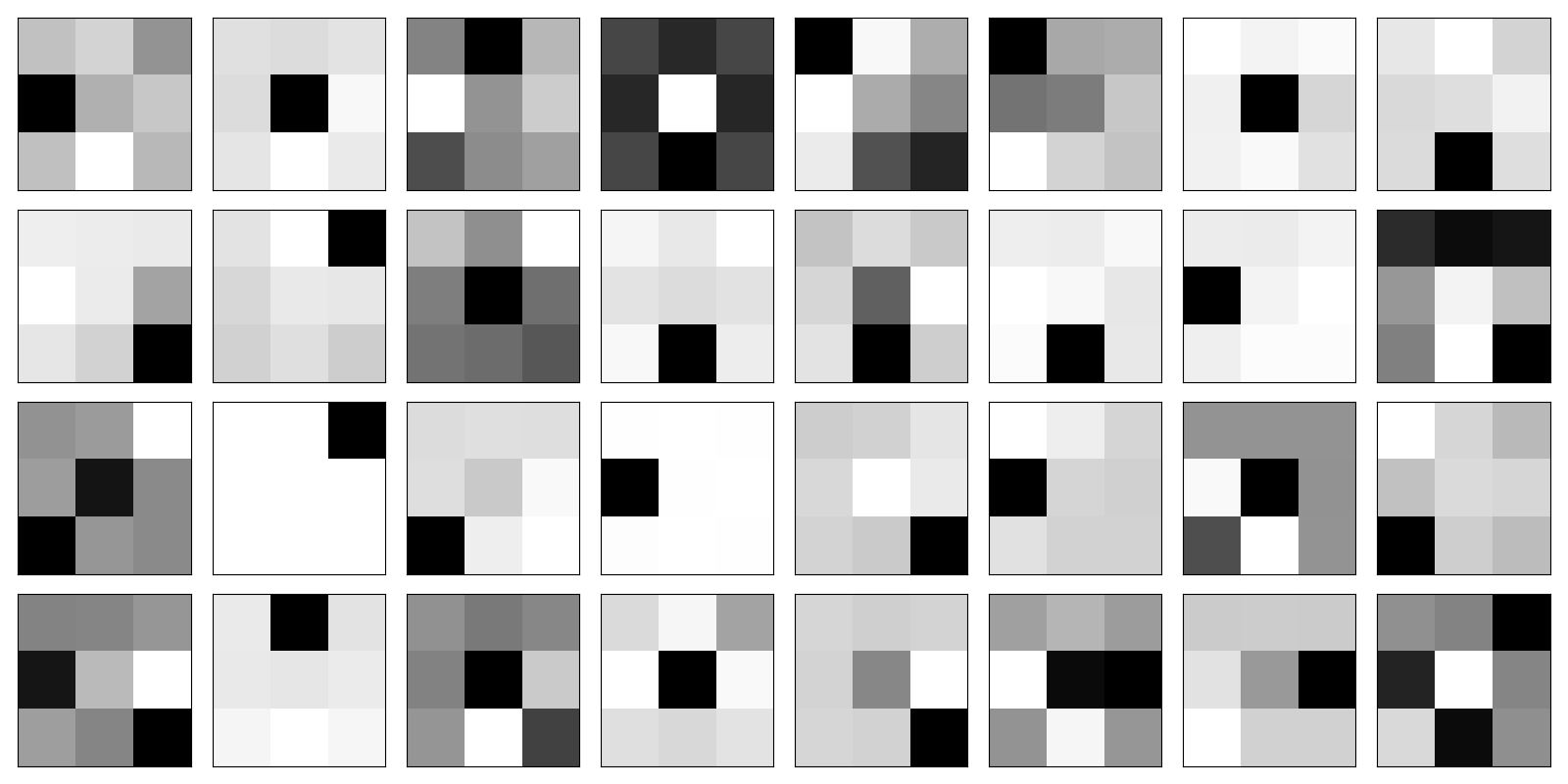}
 \caption{Example convolution kernels in the first layer of from a learned \deepCR-mask model. Gray-scale level is relative within each subplot. The network learns a diversity of filters, including one that resembles the Laplacian kernel used in \LACosmic (4th kernel from the left on the 1st row.) Note: not all trained models produce Laplacian-looking kernels.}
 \label{fig:filters}
\end{figure*}

We set the initial learning rate to 0.005 and 0.01 for \deepCR-mask and \deepCR-inpaint respectively, and use the ADAM optimizer \citep{kingma_adam:_2015}. Learning rate decays by a factor of 0.1 automatically whenever validation loss does not improve by 0.1\% for 4 epochs. We stop training once validation loss does not improve after two rounds of learning rate decay. Both \deepCR-mask variants converged within 60 epochs of training, while \deepCR-inpaint models took a longer 220-epoch training. Each epoch of training took less than 1 minute on 4 Nvidia Titan X GPUs. Figure \ref{fig:filters} shows a visualization of filters in the first convolution layer of \deepCR-mask.

\end{document}